\documentclass{aa}  

\usepackage{graphicx}
\usepackage{color,soul}
\usepackage{txfonts}
\usepackage{xcolor}
\usepackage{textcomp}
\usepackage{adjustbox}
\usepackage{subfig}
\newcommand{\qdist}[1]{\ifmmode\langle#1\rangle\else\textlangle#1\textrangle\fi}
\begin{document}

   \title{The search for extratidal star candidates around Galactic globular clusters NGC 2808, NGC 6266, and NGC 6397 with {\textit{Gaia}} DR2 astrometry}

   \author{Richa Kundu
          \inst{\ref{inst1a},\ref{inst1}} \thanks{Richa.Kundu@eso.org, richakundu92@gmail.com}
          \and
          Camila Navarrete
          \inst{\ref{inst1a},\ref{inst2}} \thanks{Camila.Navarrete@eso.org}
          \and
          Jos\'{e} G. Fern\'{a}ndez-Trincado \thanks{Jose.Fernandez@uda.cl}
          \inst{\ref{inst3}}
          \and
          Dante Minniti
          \inst{\ref{inst6},\ref{inst7}}
          \and
          Harinder P. Singh
          \inst{\ref{inst1}}
          \and
          Luca Sbordone
          \inst{\ref{inst1a}}
          \and
          Andr\'{e}s E. Piatti
          \inst{\ref{inst8},\ref{inst9}}
          \and
          C\'{e}line Reyl\'{e}
          \inst{\ref{inst9a}}
          }

   \institute{European Southern Observatory, Alonso de C\'{o}rdova 3107, 7630000 Vitacura, Santiago, Chile.
              \label{inst1a}
         \and
             Department of Physics and Astrophysics, University of Delhi, Delhi-110007, India.
              \label{inst1}
         \and             
             Millennium Institute of Astrophysics, Av. Vicu\~{n}a Mackenna 4860, 782-0436 Macul, Santiago, Chile.
              \label{inst2}
         \and
             Instituto de Astronom\'ia y Ciencias Planetarias, Universidad de Atacama, Copayapu 485, Copiap\'o, Chile.
              \label{inst3}
         \and
             Departamento de Ciencias Fisicas, Facultad de Ciencias Exactas, Universidad Andres Bello, Av. Fernandez Concha 700, Las Condes, Santiago, Chile.
              \label{inst6}
         \and
             Vatican Observatory, V00120 Vatican City State, Italy.
              \label{inst7}
        \and
            Instituto Interdisciplinario de Ciencias B\'asicas (ICB), CONICET-UNCUYO, Padre J. Contreras 1300, M5502JMA, Mendoza, Argentina.
              \label{inst8}
        \and 
            Consejo Nacional de Investigaciones Cient\'{\i}ficas y T\'ecnicas (CONICET), Godoy Cruz 2290, C1425FQB,  Buenos Aires, Argentina.
              \label{inst9}
        \and
            Institut Utinam, CNRS UMR 6213, Universit\'e Bourgogne-Franche-Comt\'e, OSU THETA Franche-Comt\'e, Observatoire de Besan\c{c}on, BP 1615, 25010 Besan\c{c}on Cedex, France.
              \label{inst9a}
             }

   \date{Received xxxx; accepted yyyy}

% \abstract{}{}{}{}{} 
% 5 {} token are mandatory
 
  \abstract
  % context heading (optional)
  % {} leave it empty if necessary  
   {Extratidal stars are stellar bodies that end up outside the tidal radius of a cluster as  a result of internal processes or external forces acting upon it. 
The presence and spatial distribution of these stars can give us insights into the past evolution of a cluster inside our Galaxy.}
  % aims heading (mandatory)
   {Previous works suggest that globular clusters, when explored in detail, show evidence of extratidal stars. We aim to search for possible extratidal stars in the Galactic globular clusters NGC 6397, NGC 2808, and NGC 6266 using the photometry and proper motion measurements from {\textit{Gaia}} DR2 database \citep{gaiadr2}.}
  % methods heading (mandatory)
   {The extratidal stars for the clusters were selected on the basis  of: their distance from the cluster center, similarity in their proper motions to the cluster population, and their position on the color-magnitude diagram of the clusters. Each cluster was explored in an annulus disk from the tidal radius up to five times the tidal radii. The significance level of the number of selected extratidal stars was determined on the basis of the distribution of Milky Way stars
according to the Besan\c{c}on Galaxy model and {\textit{Gaia}} data. To understand the observed extratidal features, the orbits of the clusters were also determined using \texttt{GravPot16}.}
  % results heading (mandatory)
   {Finally, 120, 126, and 107 extratidal candidate stars were found lying outside the tidal radius of the globular clusters NGC 6397, NGC 2808, and NGC 6266, respectively. 70\%, 25.4\%, and 72.9\% of the extratidal stars found are located outside the Jacobi radius of NGC 6397, NGC 2808, and NGC 6266, respectively. The spatial distribution of the extratidal stars belonging to NGC 6397 appears S-like, extending along the curved leading and trailing arms. NGC 2808 has an overdensity of stars in the trailing part of the cluster and NGC 6266 seems to have overdensities of extratidal stars in its eastern and northern sides.}
  % conclusions heading (optional), leave it empty if necessary 
   {Proper motions and color-magnitude diagrams can be used to identify extratidal candidate stars around GCs. Nonetheless, depending on how different the kinematics and stellar populations of a cluster are compared to the Milky Way field, the fraction of contamination can be larger. All three clusters are found to have extratidal stars outside their tidal radii. {For NGC 6397 and NGC 2808, these stars} may be the result of a combined effect of the disc shocks and tidal disruptions. For NGC 6266, the distribution of extratidal stars is symmetrical around it, most likely indicating that the cluster has an extended stellar envelope.}

\titlerunning{Extra-tidal stars in NGC 2808, NGC 6266 \& NGC 6397}
\authorrunning{Kundu et al.}

   \keywords{(Galaxy:) globular clusters: general, (Galaxy:) globular clusters: individual: NGC 6397, (Galaxy:) globular clusters: individual: NGC 6266, (Galaxy:) globular clusters: individual: NGC 2808}

   \maketitle
%
%________________________________________________________________

\section{Introduction}

Galactic globular clusters (GCs) are key  to improving our understanding the formation and evolution of our Galaxy. In most GCs, stars are formed over a very short time span and with very similar compositions in most chemical elements \citep{Bastian18}, and they are all at the same distance from our planet. These qualities make Galactic GCs apt for the study of stellar evolution, as are their stars, which are in a range of different evolutionary phases across a single system. Studying the extratidal region\footnote{In this manuscript, we consider a star to be extratidal when it is located beyond the observational tidal radius ($r_t$). Generally, $r_t$ is estimated as the radius at which the surface density profile of the cluster population drops to zero density \citep[either with a King, Wilson, or LIMEPY model, see e.g.,][and references therein]{deboer19}.}, that is, the region which is out of the $r_t$ of a given cluster, can give a better insight into the different internal processes acting on the cluster itself, including stellar evolution, gas expulsion, and two-body relaxation \citep{Geyer2000}, as well as the forces acting on the cluster which may strip stars from it. These stars can be considered as "potential escapers" as they are still bound to the cluster if they are inside the Jacobi radius ($r_J$), corresponding to the boundary distance from the cluster center at which a star is still bound. For more details, see, for example, \citet{Fukushige2000, Baumgardt10, Kupper10,  julio11, Claydon10} and references therein. Disk or bulge shocking, tidal disruptions, relaxation, and dynamical friction may produce these potential escapers, together or independently, which may lead to their leaving the influence of the gravitational potential of the cluster and forming tidal tails and halos around these systems \citep{Leon00, Odenkirchen2001, Moreno14, Hozumi14, Balbinot18}. However, these effects have been proven to have a minor impact in the morphology of tidal tails \citep[see, e.g.,][]{BaumgardMakino}.
\\

Observational evidence of extratidal stars around GCs have been found in the form of extended tidal tails \citep[see e.g.,][]{Grillmair2006, ostholt10, sollima11, balbinot11, myeong17, camila17, Bonaca20} and the asymmetric distribution of stars in the immediate outskirts of some clusters \citep[e.g., ][]{jose16, julio17, kundu19a}, as well as through the chemo-dynamic detection of stars that have shown CN anomalies on the outskirts of some GCs \citep{Hankw20} and as debris stars throughout the inner and outer stellar halo \citep{costa2008, majewski12, jose15a, jose15,Fernandez-Trincado2016b,jose19c,jose19halo}. However, it is still not clear why some GCs show signs of extratidal material, but scarcely any extratidal stars or tidal tails, while others lack any evidence for such structures \citep[see, e.g.,][]{piatti20}.
\\

A complete census of potential extratidal structures in the outskirts of GCs could help to  provide a better understanding of the nature of the inner and outer stellar halo. In general, the light-element abundances of cluster stars are key to revealing the origin of halo field stars with unique chemical signatures throughout the MW \citep[this is the so-called "chemical-tagging" method, see e.g.,][]{Martell10,jose19c,jose19halo,jose2020a,jose2020b}. In this sense, the \textit{Gaia} data release 2 \citep[][hereafter, DR2]{gaiadr2} astrometry allows us to achieve a homogeneous exploration in the immediate vicinity around GCs to probe the existence or absence of potential extratidal stars for future spectroscopic follow-ups.
\\

In this work, we take advantage of the \textit{Gaia} DR2 mission to examine the outermost regions of three GCs buried in different Galactic environments, that is, NGC 6397, NGC 2808 and NGC 6266 (M 62). The exquisite data from {\textit{Gaia}} DR2 allows us to homogeneously improve and increase the number of dimensions of the parameter space to select potential cluster members in the outermost regions of these GCs, which were chosen for a number of reasons. First, NGC 6397 has been dynamically classified as a main-disk GC \citep{Massari19} even though it has a mean metallicity of [Fe/H]= -1.88 \citep{Correnti2016, Meszaros2020}, which is largely offset from the metallicity of disk field stars. 
\\

NGC 6266 has been classified as a main-bulge cluster \citep{Massari19}, with a mean metallicity of -1.29 dex \citep{Correnti2018}, which is similar to that of Si-/N-/Al-rich stars that were recently found in the bulge region \citep{jose2020a,jose2020b}, which could be part of an ancient GC population that formed the bulge \citep[although their origin is still under debate, see e.g.,][]{bekki19}. NGC 2808 is known to be a massive cluster, hosting several stellar populations \citep{Piotto07, Milone15a, Latour19} and has been recently associated with the accreted Gaia-Sausage-Enceladus \citep{Belokurov2018, Haywood2018, Helmi2018, Koppelman2018, Myeong2018} dwarf galaxy, which is supposed to dominate the inner halo stellar population. 
\\

With regard to the analysis of possible extratidal stars, these three clusters are easily accessible in terms of distance:\ NGC 6397 is the second-closest GC, located at a distance of 2.3 kpc \citep[2010 edition of ][hereafter, H96]{Harris96}, only after NGC 6121 (M4, at 2.2 kpc). Both NGC 6266 (6.8 kpc) and NGC 2808 (9.6 kpc) are closer than 10 kpc (H96). All these clusters are located in regions with high stellar density and relatively high reddening values of E(B-V)= 0.22, 0.21, and 0.47 mag for NGC 6397 \citep{Correnti2018}, NGC 2808 \citep{Correnti2016} and NGC 6266 (H96), respectively. These conditions could affect the reliability of potential extratidal star detections if only photometry is considered, but more reliable candidates can be obtained using proper motions (PMs) from \textit{Gaia} DR2 \citep[see, e.g.,][]{Piatti20a}. 
\\

It is worth mentioning that some evidence of extratidal material has been claimed in the literature specifically for NGC 6266 based on near-infrared photometry, without taking into account PMs \citep{Chun15}, and using RR Lyrae stars \citep{dante18}, while evidence for extratidal stars around NGC 2808 was found by \citet{julio17} based on deep photometry, without including PMs. Previously, \citet{Leon00} detected tidal tails for NGC 6397.
\\

Our previous works suggest that several GCs, when explored in detail, show some evidence for extratidal stars \citep{jose15, jose15a, jose16, camila17, dante18, kundu19, kundu19a, Piatti20a}. Here, we exploit the superb Gaia DR2 data-set in order to address the issue of the existence of the extratidal features around three GCs, using both photometry and astrometry and their intrinsic limitations. The candidate extratidal stars identified in this study could be investigated in follow-up spectroscopic campaigns, such as the SDSS-V Pioneering Panoptic Spectroscopy survey \citep[see, e.g.,][]{Kollmeier2017}, to fully characterize them, both chemically and dynamically.
\\

This paper is organised as follows: Section 2 discusses the criteria used to select the extratidal stars based on their position, PMs, and color-magnitude diagrams (CMDs) of the clusters. In Section 3, we carry out a backwards integration of the orbits of the clusters and derive updated orbital parameters and membership values to the disk, bulge, and halo Galaxy components. Finally, in Section 4, we discuss our results and present our  main conclusions. \\

\section{Selecting extratidal star candidates}
\label{sec2}

In this work, we study the outer region of the Galactic GCs NGC 6397, NGC 6266, and NGC 2808 using the {\textit{Gaia}} DR2 catalog. We adopted the same procedure followed by \citet{kundu19} to clean the sample, thereby eliminating any contamination due to data processing artifacts or spurious measurements, as suggested by the {\textit{Gaia}} collaboration \citep[for details refers to Section 2 in][]{kundu19}.
\\

\begin{figure}
\subfloat {\includegraphics[width = 3in]{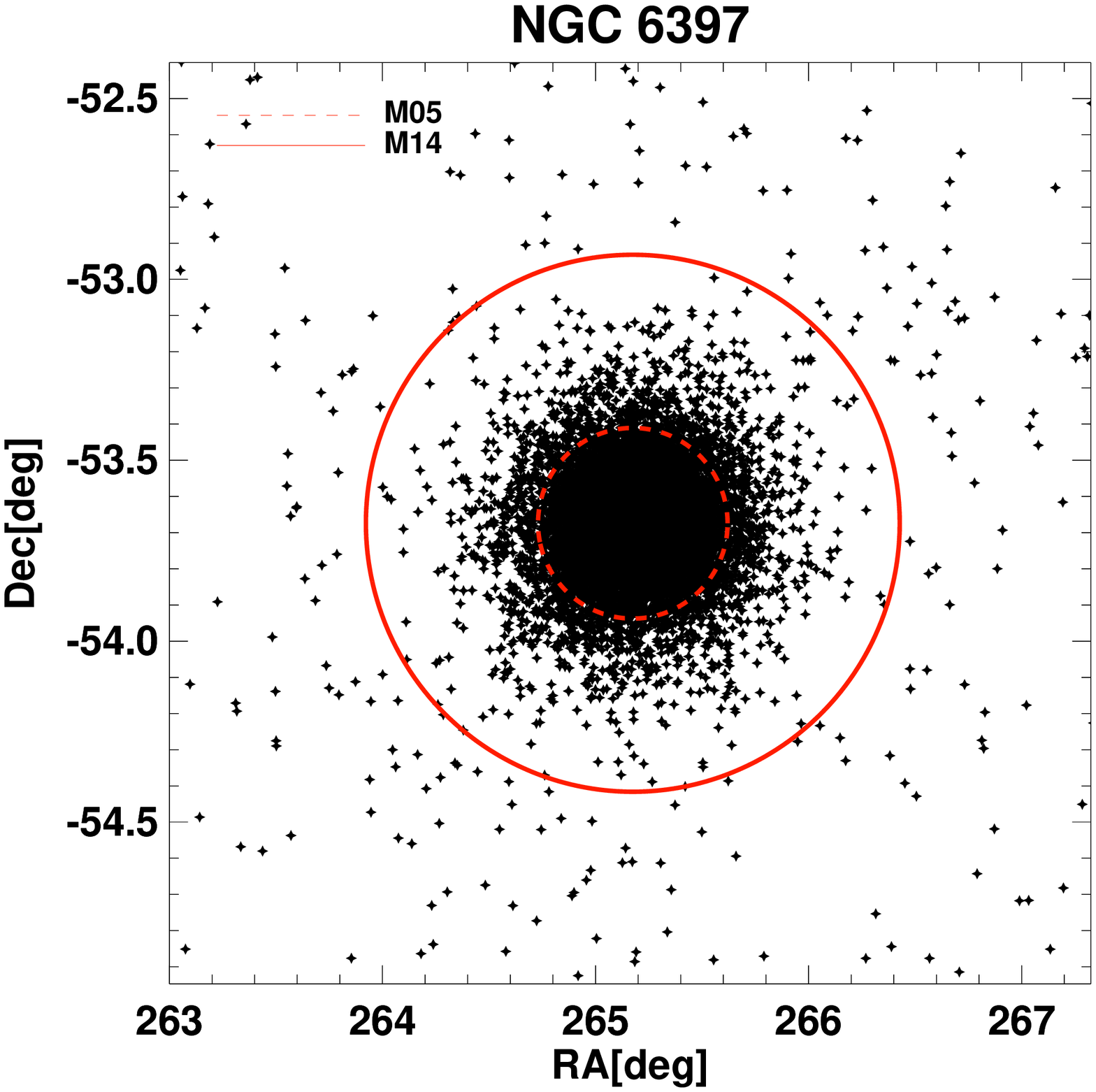}}\\
\subfloat {\includegraphics[width = 3in]{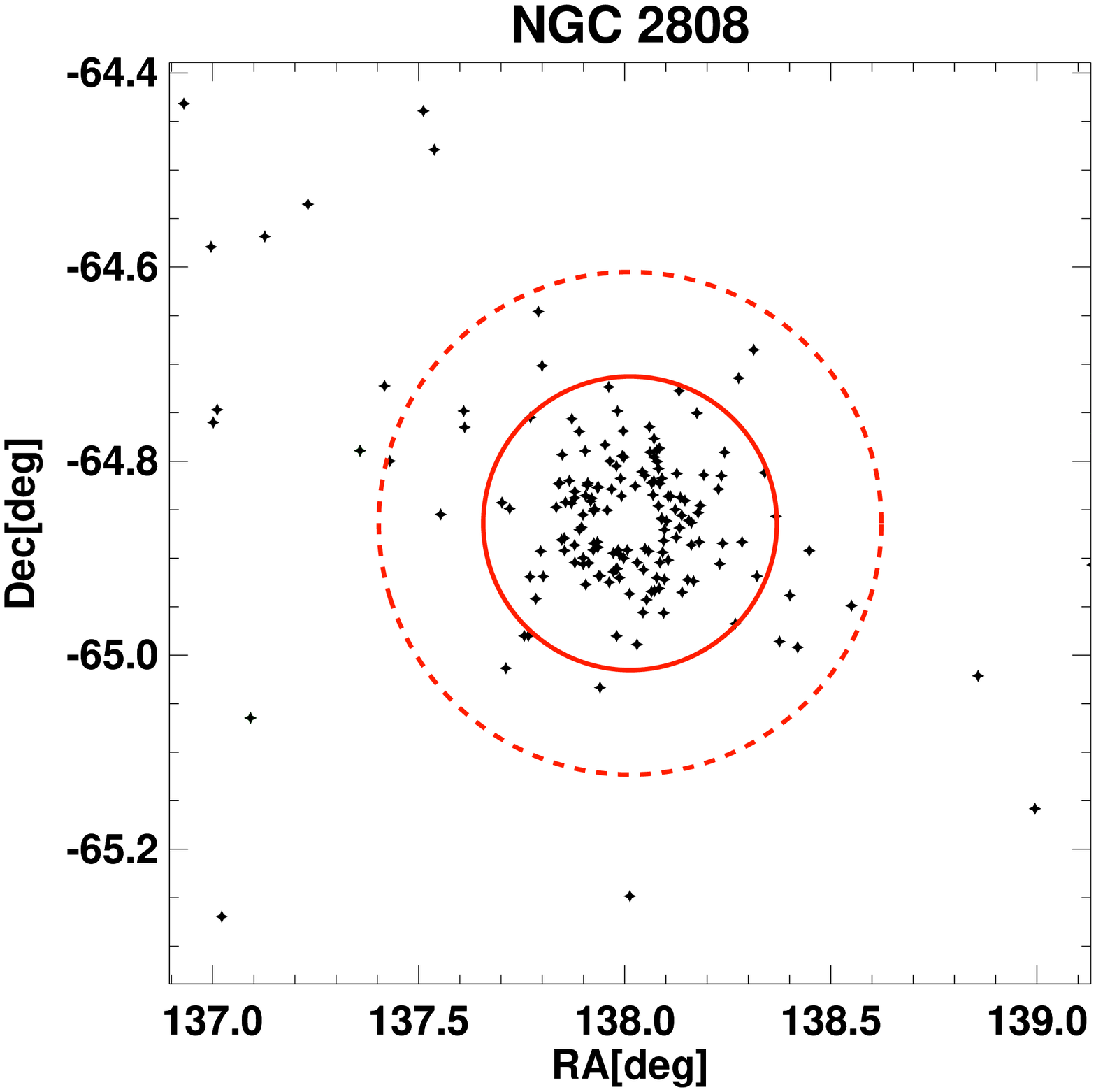}}\\
\subfloat {\includegraphics[width = 3in]{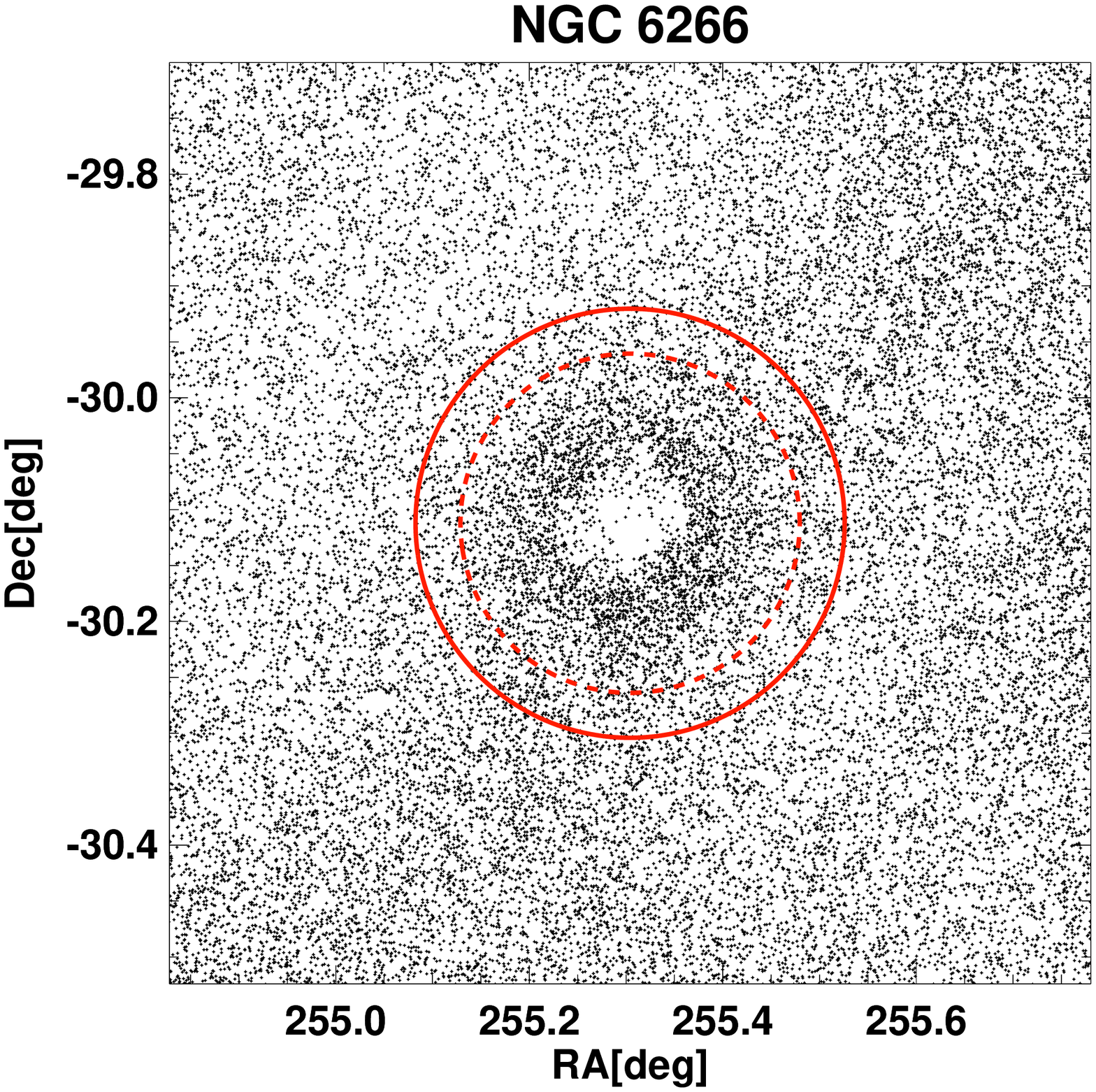}}
\caption{Black dots are the {\textit{Gaia}} DR2 stars whose PM is similar to the mean PM of the cluster within 3 sigma \citep[as listed in][]{GC}. The solid and dashed red circles correspond to the $r_t$ from \citet{Moreno14} and \citet{mackey05}, respectively.}
\label{fig:TR}
\end{figure}

We began our analysis by selecting the $r_t$ values of the clusters to adopt in this study. In order to do so, we first selected the stars with PMs within three sigma of the mean value \citep[as listed in][]{GC} in a region covering two degrees around the cluster center. Figure~\ref{fig:TR} shows the spatial distribution of the selected stars along with the values for the $r_t$ from \citet{mackey05} and \citet{Moreno14}. We can see from Figure~\ref{fig:TR} that for NGC 6397, the $r_t$ from \citet{mackey05} is underestimated, hence, we adopted the value for $r_t$ reported in \citet{Moreno14} for this analysis. For the other two clusters, the $r_t$ values provided by \citet{mackey05} seem to bound all the cluster stars and, therefore, we chose these values. The adopted values for r$_t$ are the following: 15.55 arcmin for NGC 2808, 44.53 arcmin for NGC 6397, and 8.97 arcmin for NGC 6266.
\\

Once we adopted a $r_t$ for each cluster, we estimated the mean and the intrinsic dispersion of the PM distribution of each cluster. It is worth mentioning that the uncertainties listed in \citet{GC} for the mean PMs take into account the statistical and systematic uncertainties but they do not represent the intrinsic dispersion of the PM distribution, which can be up to ten times larger. Therefore, to consider the intrinsic dispersion in the PM distribution, we selected {\textit{Gaia}} data for stars up to the $r_t$ for the three clusters. A Gaussian mixture model consisting of two Gaussians (one for the cluster and one for the field stellar populations) was fitted to $\mu_{\alpha}\cos{\delta}$ and $\mu_{\delta}$ independently, without the need to take into account correlated errors between these two quantities. Gaussian mixture models for the PM distribution of cluster and field stars have been used to measure mean proper motions of globular clusters \citep[see e.g.,][]{dana10,Baumgardt2019}, including the errors on the measurements. It is beyond the scope of this work to model the exact shape of the PM distribution, convolving it with the error function, of each cluster but to have an estimate of the intrinsic dispersion in order to select extratidal star candidates. The results from the fit were adopted as the mean PM (center of the Gaussian) and the intrinsic dispersion (one sigma) for the distribution of $\mu_{\alpha}\cos{\delta}$ and $\mu_{\delta}$ of the cluster population. Our PM values match the reported values in \citet{GC}, within the errors. A proper fit of the PM distribution of the cluster stars including PM in both directions (RA and Dec at the same time) is beyond the scope of this paper, although it has been used in the literature to estimate the membership probability and uncover, in combination with parallaxes and CMDs, the tidal tails in several clusters. For an example, see \citet{Sollima20}.
\\

Once we had both the $r_t$ and the PMs for the clusters, we selected the extratidal stars based on three main criteria: the position of the stars with respect to the cluster centers, PMs of the clusters and the stars, and the position of the stars on the CMDs. First, we selected the stars which lie on an annular disk centered on the cluster, having as inner and outer radii of one and five times the $r_t$ of each cluster. Next, to remove field stars based on PMs, we selected only those stars whose PMs match the PM of the cluster, within the combined error bar of the pair GC-star, that is, \{($\mu_{\alpha}\cos{\delta}\pm\sigma$), ($\mu_{\delta}\pm\sigma$)\}$^{star}$ $\lesssim$ \{($\mu_{\alpha}\cos{\delta}\pm\sigma$), ($\mu_{\delta}\pm\sigma$)\}$^{cluster}$, with $\sigma^{star}$ as the error in the PM of the star and $\sigma^{cluster}$ the dispersion in the PM distribution of the cluster. Finally, we selected stars based on the PARSEC isochrones\footnote{http://stev.oapd.inaf.it/cmd} \citep{Bressan12,Marigo17} for the clusters. We selected only those stars that lie within a 0.01 magnitude/color ratio away from the isochrone. 
\\

\begin{figure}
\subfloat {\includegraphics[width = 3in]{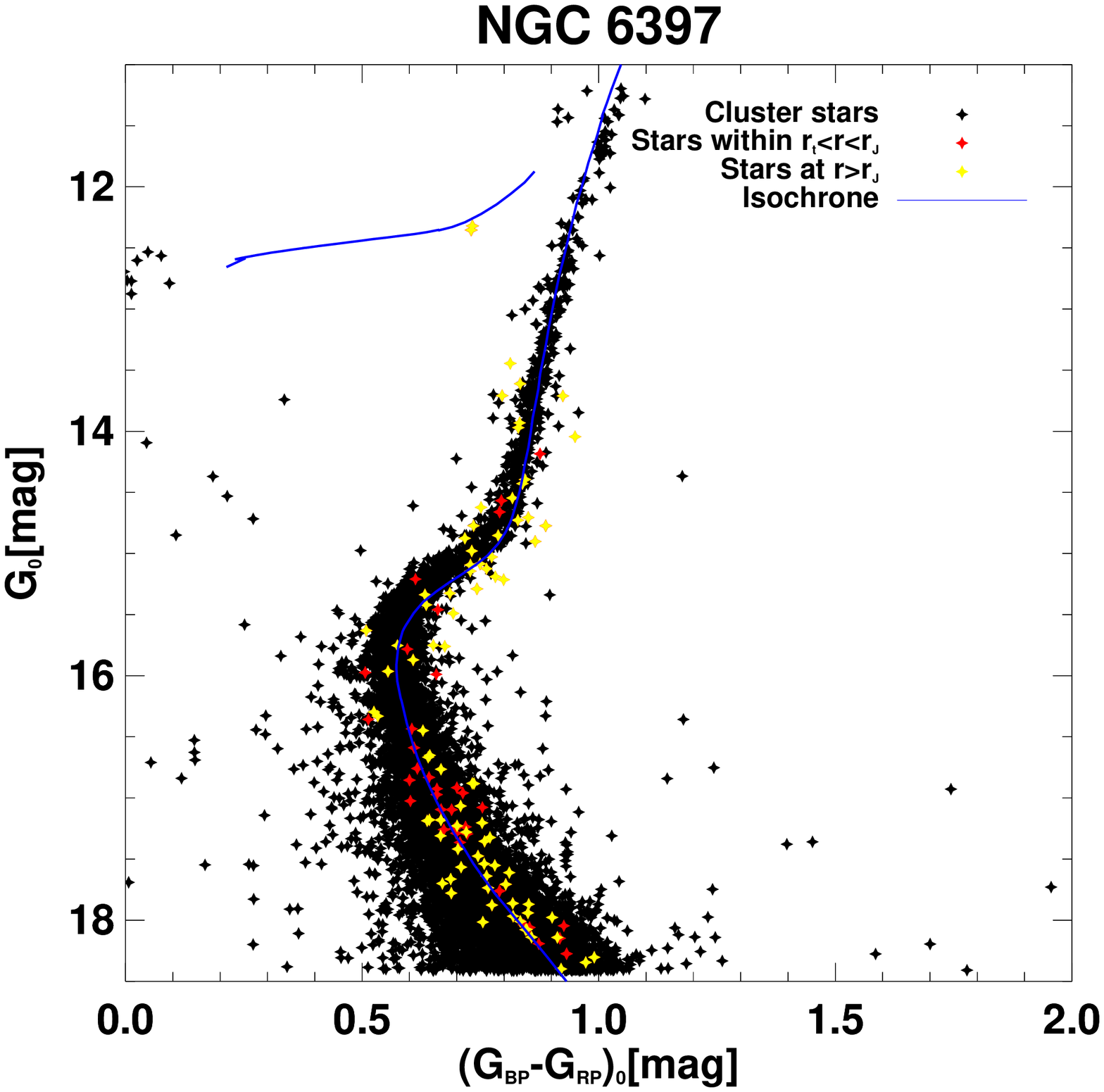}}\\
\subfloat {\includegraphics[width = 3in]{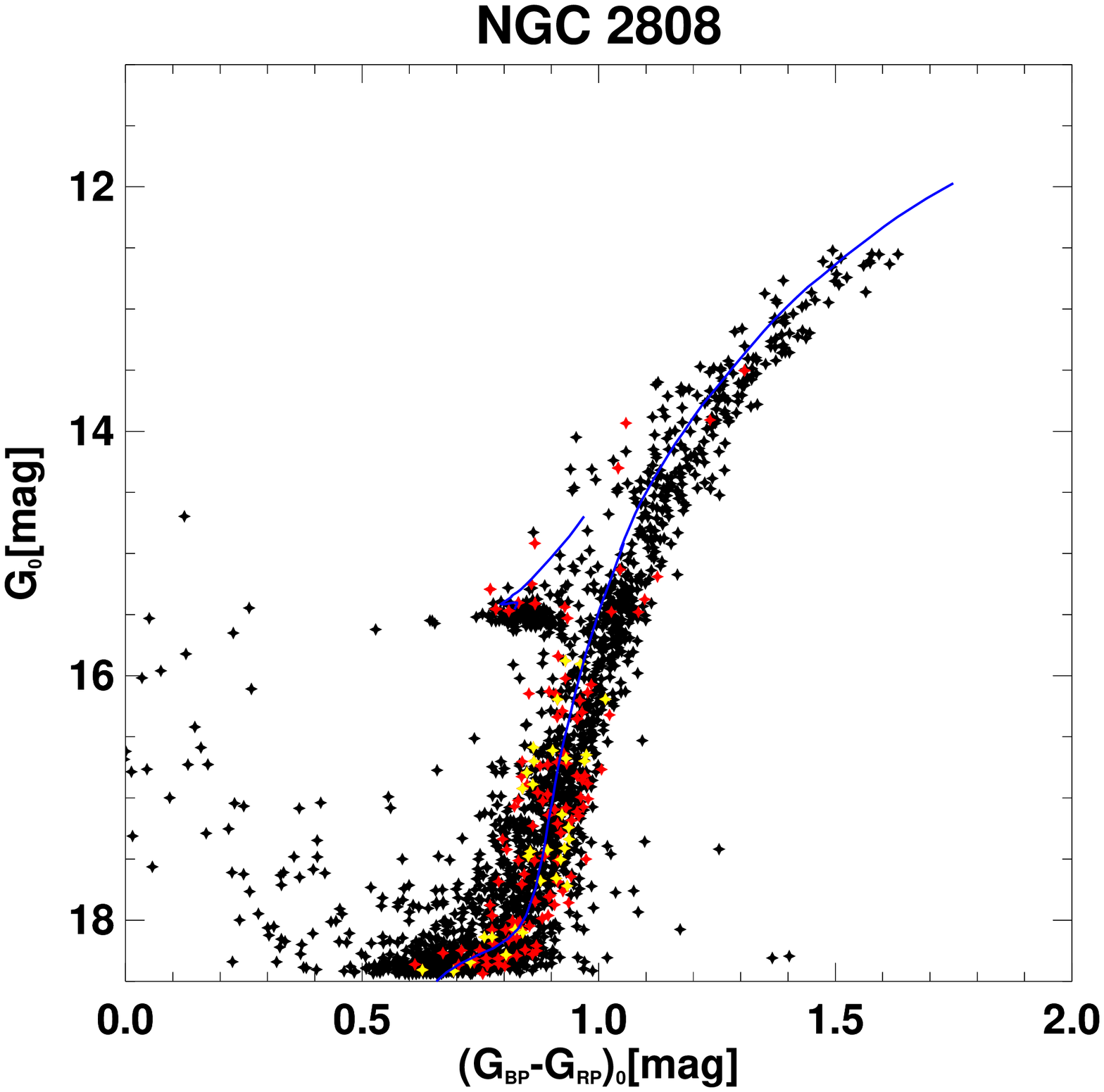}}\\
\subfloat {\includegraphics[width = 3in]{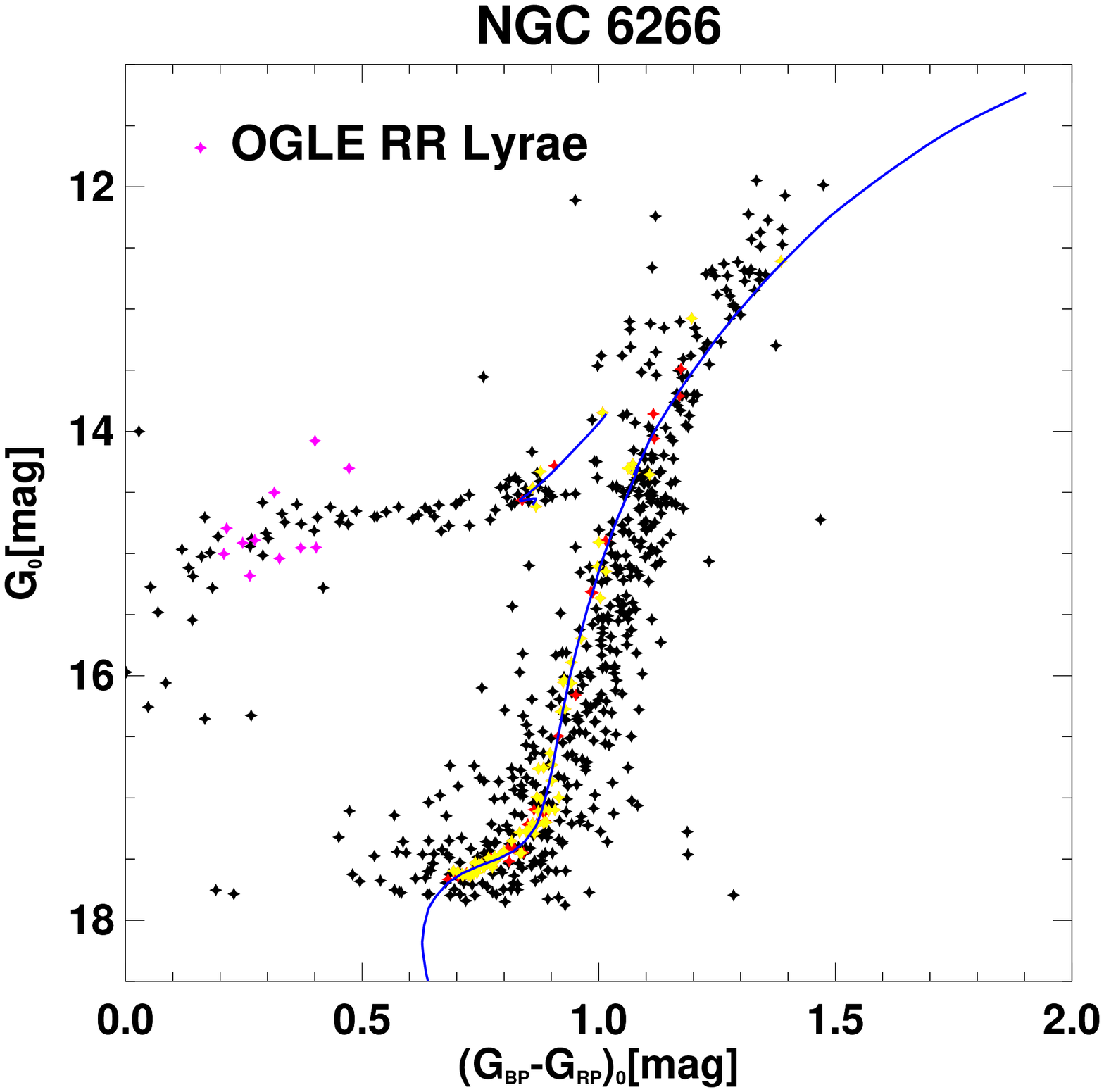}}
\caption{De-reddened CMDs of the clusters in {\textit{Gaia}} DR2 bands. Cluster stars (stars within the $r_t$ of the cluster) are shown with black dots and selected extratidal stars within and outside the cluster r$_J$ are shown in red and yellow, respectively. For NGC 6266, OGLE RR Lyrae stars are shown in pink (see Section~\ref{RRL}).}
\label{fig:f1}
\end{figure}

\begin{figure}
\subfloat {\includegraphics[width = 3in]{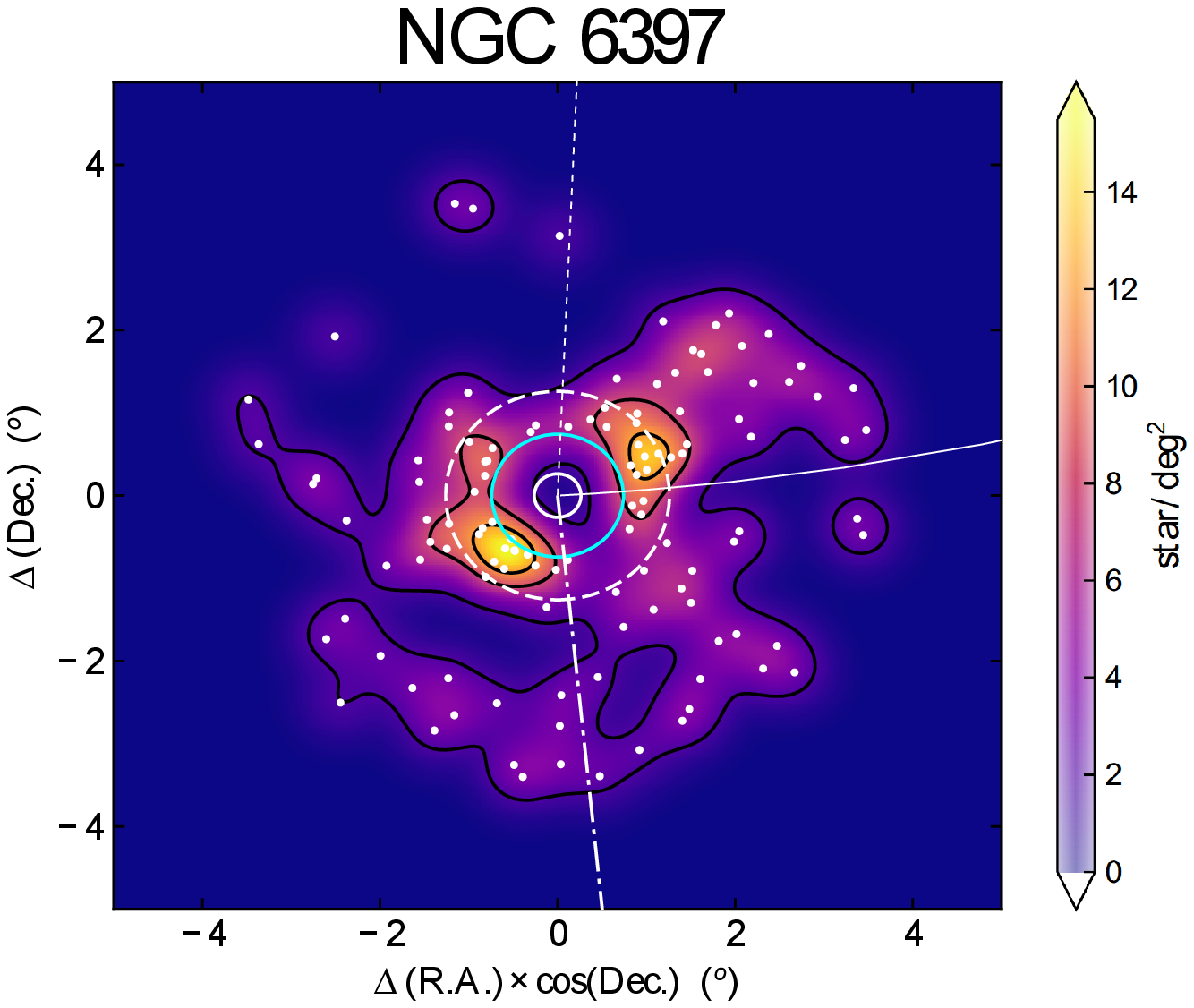}}\\
\subfloat {\includegraphics[width = 3in]{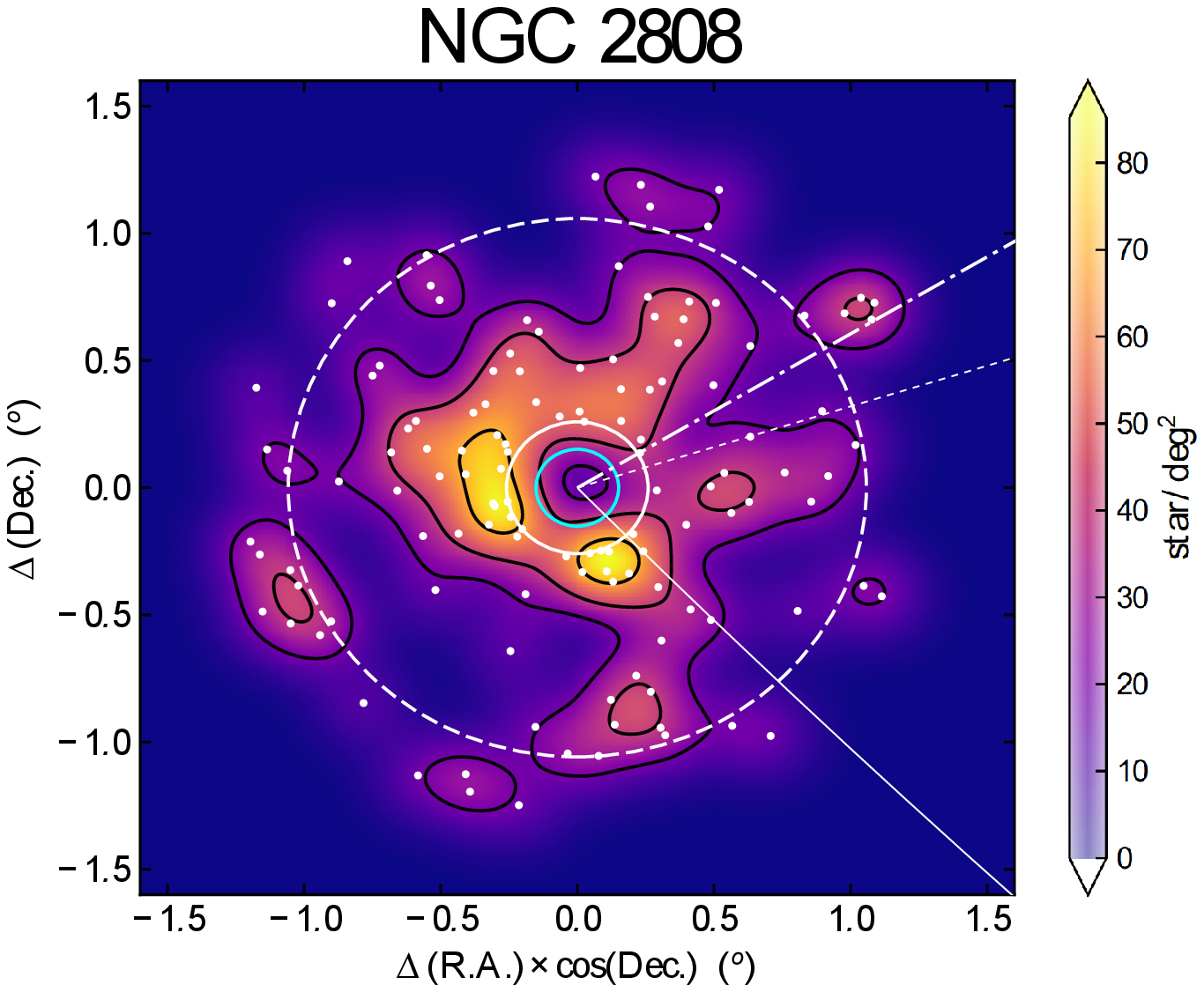}}\\
\subfloat {\includegraphics[width = 3in]{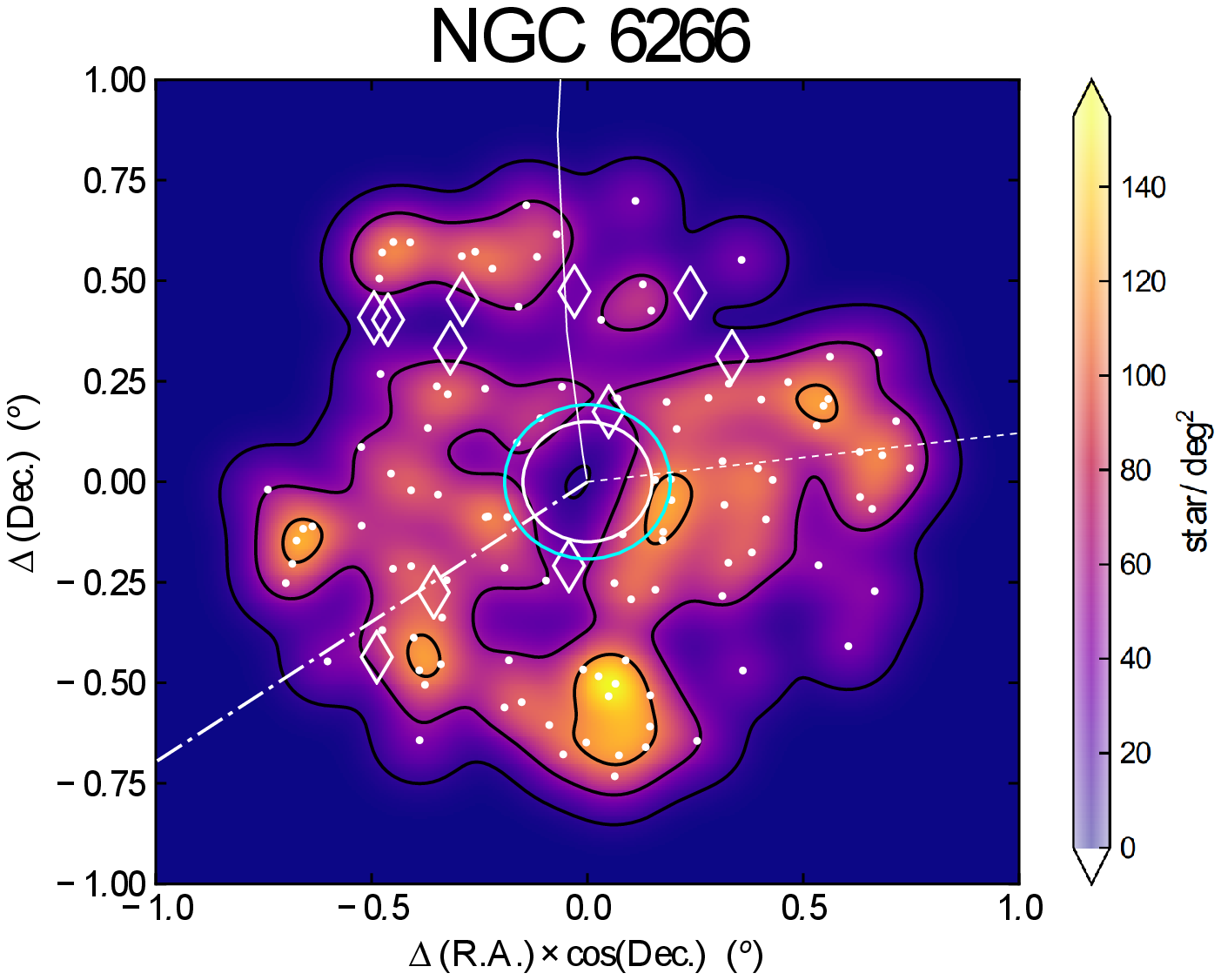}}
\caption{Stellar density maps built from extratidal star candidates that occupy the CMD in Figure \ref{fig:f1}. The white and cyan circle centered on each cluster indicates the $r_{t}$ from \citet{mackey05} and from \citet{Moreno14}, respectively. While the dashed circle indicate the $r_{J}$ from \citet{deboer19}. The white lines indicate the directions of the cluster PM (dash-dot), the Galactic center (dotted), and the cluster orbit (solid) computed with the \texttt{GravPot16} code. Diamond symbols indicate the RR Lyrae stars.}
\label{fig:f2}
\end{figure}

The number of stars selected at each step along with the cleaning and after, along with each selection criterion, is provided in Table~\ref{Tabla1}. The first five lines list the cleaning criteria and after that selection criteria are provided. The de-reddened CMDs of the clusters (member stars as black dots) along with the isochrones (blue) and selected stars (red for stars within the $r_J$ and yellow for stars out of the $r_J$) are shown in Figure~\ref{fig:f1}. All {\textit{Gaia}} magnitudes were de-reddened using the individual E(B$-$V) values from \citet{Schlafly11}\footnote{\url{https://irsa.ipac.caltech.edu/applications/DUST/}} dust maps and the {\textit{Gaia}} extinction coefficients provided by \citet{gaiadr2}. \citet{Alonso12} showed that NGC 6266 suffers from considerable differential reddening and, therefore, we used their reddening map to de-redden the regions around the cluster where we searched for extratidal stars. However, the data are available only for the stars inside the $r_t$, hence we used \citet{Schlafly11} maps to de-redden the extratidal stars around the cluster. The stars which lie within the $r_t$ of the clusters and have PMs similar to the clusters were selected as cluster stars for the CMDs. Figure~\ref{fig:f2} shows the spatial distribution of the selected stars (white dots), along with the $r_t$ (white solid circle is from \citealt{mackey05} and the cyan solid circle is from \citealt{Moreno14}) and $r_J$ \citep[taken from][shown as a white dashed circle]{deboer19} of each cluster. The dotted-dashed line in Figure~\ref{fig:f2} shows the direction of the mean PM of the cluster and the dotted line points towards the Galactic center. These plots were made using the kernel density estimator (KDE) routine in AstroML \citep{VanderPlas12}, using a grid of 400 pixels in each direction. The bandwidth of the Gaussian KDE used was 24.0 arcmin for NGC 6397, 8.4 arcmin for NGC 2808 and 4.2 arcmin for NGC 6266. The contours mark the levels with more than [3,5,9], [10,35,75], and [10,50,110] stars per square degree for NGC 6397, NGC 2808 and NGC 6266, respectively. The adopted parameters for the selection process and our results for each cluster are discussed in the following subsections.
\\

%%%%%%%%%%%%%%%%%%%%%%%%%%%%%%%%%%%%%%%%%%%%%%%%%%
\begin{table}
        \centering
        \caption{Number of selected stars passing each high-quality criterion and different selection cuts.}
        \label{Tabla1}
        \begin{adjustbox}{width=1.0\columnwidth,center}
        \begin{tabular}{|lr|} % four columns, alignment for each
\hline
Criteria & Number of stars \\
\hline
\hline
NGC~6397&\\
\hline
\hline
\texttt{1.) ASTROMETRIC\_GOF\_AL $<$ 3} & 8,445,370 \\
\texttt{2.) ASTROMETRIC\_EXCESS\_NOISE\_SIG $\leq$ 2.} & 8,200,243 \\
\texttt{3.) $-$0.23 $\leq$ MEAN\_VARPI\_FACTOR\_AL $\leq$ 0.32} & 8,195,763 \\     
\texttt{4.) VISIBILITY\_PERIODS\_USED > 8} & 7,768,205 \\      
\texttt{5.) G $<$ 19 mag} & 3,507,704 \\  
\texttt{6.) Between $r_{t}$ and $5\times$ $r_{t}$} &  1,789,728\\
\texttt{7.) With similar PM as the cluster} & 434 \\       
\texttt{8.) Stars in the cluster CMD} & {\bf 120} \\
\hline
\hline
NGC~2808& \\
\hline
\hline
\texttt{1.) ASTROMETRIC\_GOF\_AL $<$ 3} & 286,944 \\
\texttt{2.) ASTROMETRIC\_EXCESS\_NOISE\_SIG $\leq$ 2.} & 280,162 \\
\texttt{3.) $-$0.23 $\leq$ MEAN\_VARPI\_FACTOR\_AL $\leq$ 0.32} & 275,781 \\     
\texttt{4.) VISIBILITY\_PERIODS\_USED > 8} & 273,149 \\      
\texttt{5.) G $<$ 19 mag} & 125,699 \\ 
\texttt{6.) Between $r_{t}$ and $5\times$ $r_{t}$} &  101,782\\
\texttt{7.) With similar PM as the cluster} & 424 \\        
\texttt{8.) Stars in the cluster CMD} & {\bf 126} \\
\hline
\hline
NGC~6266 (with normal cuts)&\\
\hline
\hline
\texttt{1.) ASTROMETRIC\_GOF\_AL $<$ 3} & 7,008,462 \\
\texttt{2.) ASTROMETRIC\_EXCESS\_NOISE\_SIG $\leq$ 2.} & 6,667,043 \\
\texttt{3.) $-$0.23 $\leq$ MEAN\_VARPI\_FACTOR\_AL $\leq$ 0.32} & 6,156,844 \\     
\texttt{4.) VISIBILITY\_PERIODS\_USED > 8} & 3,636,806 \\      
\texttt{5.) G $<$ 19 mag} & 1,228,027 \\ 
\texttt{6.) Between $r_{t}$ and $5\times$ $r_{t}$} & 159,567\\
\texttt{7.) With similar PM as the cluster} & 6,784 \\       
\texttt{8.) Stars in the cluster CMD} & {\bf 2,155} \\
\hline
\hline
\hline
NGC~6266 (with stricter cuts)&\\
\hline
\hline 
\texttt{1.) Between $r_{t}$ and $5\times$ $r_{t}$} & 159,567\\
\texttt{2.) With similar PM as the cluster} & 1,729 \\       
\texttt{3.) Stars in the cluster CMD} & {\bf 107} \\
\hline
\end{tabular}
\end{adjustbox}
\end{table}
%%%%%%%%%%%%%%%%%%%%%%%%%%%%%%%%%%%%%%%%%%%%%%%%%%

\subsection{NGC 6397}

The mean PM of the cluster, as determined by the Gaussian fitting model is: $\mu_{\alpha}\cos{\delta}$= 3.302$\pm$0.540 mas yr$^{-1}$; $\mu_{\delta}$=-17.600$\pm$0.631 mas yr$^{-1}$. Here, 0.540 mas yr$^{-1}$ and 0.631 mas yr$^{-1}$ are the associated dispersions in $\mu_{\alpha}\cos{\delta}$ and $\mu_{\delta}$, respectively. The input parameters used to get the isochrone were taken from \citet{Correnti2018}, as they give a better description of the CMD than the parameters listed in H96. In particular, we consider an age of 12.6 Gyr, [Fe/H] = -1.88, distance modulus of 12.1 mag. Then, applying the different cuts explained earlier, we found 120 extratidal stars around the cluster. Out of these 120 extratidal stars, 85 of them are outside the $r_J$ \citep[75.6 arcmin, ][]{deboer19}. The top panel of Figure~\ref{fig:f1} shows the CMD of the cluster along with the selected extratidal stars. The top panel of Figure~\ref{fig:f2} shows the density map of these 120 extratidal stars (white dots).
\\

\subsection{NGC 2808}

The mean PM and one sigma dispersion of the cluster, as determined by the Gaussian fitting model, are: $\mu_{\alpha}\cos{\delta}$= 1.087$\pm$0.620 mas yr$^{-1}$; $\mu_{\delta}$=0.248$\pm$0.503 mas yr$^{-1}$. The parameters used to download the isochrone are taken from \citet{Correnti2016}: 11.2 Gyr, [Fe/H] = -1.23, distance modulus of 15.09 mag. NGC 2808 is the most distant cluster out of the three. Hence, we also used the individual stellar parallaxes provided by {\textit{Gaia}} DR2 to reject the obvious foreground stars. In particular, we rejected all the stars whose parallax is larger than 0.5 mas (i.e., stars at distances less than 2 kpc). In the final selection, there are 126 extratidal stars and their position on the CMD of the cluster and spatial distribution are shown in the middle panels of Figure~\ref{fig:f1} and \ref{fig:f2}, respectively. Out of these 126 stars, 32 extratidal stars lie outside the r$_J$ of the cluster \citep[63.5 arcmin, ][]{deboer19}.
\\

\subsection{NGC 6266}

The mean PM of the cluster, as determined by the Gaussian fitting model, are: $\mu_{\alpha}\cos{\delta}$= -5.047$\pm$0.674 mas yr$^{-1}$; $\mu_{\delta}$=-3.021$\pm$0.566 mas yr$^{-1}$. The isochrone for the cluster is downloaded adopting the metallicity [Fe/H] = -1.02, age = 11.78 Gyr \citep{Forbes10} and distance modulus of 14.2 mag, a slightly higher value than in H96 (15.64 mag) to fit the isochrone to the horizontal-branch level of cluster. Based on the PM and cuts in the CMD, 2155 extratidal stars were selected. This cluster is located in a high-density region and our selection suffers from a high fraction of contaminants (see next Section). The CMD and spatial map for the extra-tidal stars found, after stricter cuts (see Section 2.5) are shown in the bottom panels of Figures~\ref{fig:f1} and ~\ref{fig:f2}, respectively. 
\\

The tables containing final set of selected candidates are only available in electronic form at the CDS via anonymous ftp to cdsarc.u-strasbg.fr (130.79.128.5) or via \url{http://cdsweb.u-strasbg.fr/cgi-bin/qcat?J/A+A/}.

\subsection{Significance of extratidal stars}
\label{sec_sig}

For each cluster, the significance of the potential extra-tidal star candidates was examined. To this purpose, we made use of the updated version of the Besan\c{c}on Galaxy model\footnote{\url{www.model.obs-besancon.fr}} \citep[hereafter BGM,][]{Robin2003} in order to get a rough estimation of the expected Galactic contamination along the regions examined in this study.
\\

The BGM makes use of the population synthesis approach that simulates observations of the sky with errors and biases. It is based on a scenario for Galaxy formation and evolution that reflects our present knowledge about the Milky Way (MW). Four stellar populations are considered in the model: a thin disk, a thick disk, a bar, and a halo, with each stellar population having a specific density distribution. Our simulations were done using the revised scheme of BGM \citep{Czekaj2014} where the stellar content of each population is modelled through an initial mass function, a star formation history, and follows evolutionary tracks \citep[revised in][]{Lagarde2017}. The resulting astrophysical parameters are used to compute their observational properties, using atmosphere models, and assuming a 3D extinction map computed from \citet{Marshall2006} and \citet{Lallement2019}. It includes the simulation of binarity, while the merging is done assuming the 0.4 arcsec spatial resolution of {\textit{Gaia}} DR2. A dynamical model is used to compute radial velocities and PMs, as described in \cite{Robin2017}.
\\

We roughly estimated the number of MW stars (false positives) passing our criteria, which were then compared with our potential extratidal star candidates given in Section~\ref{sec2}. In this way, we get 48 field stars for NGC 2808, 24 for NGC 6397, and 5184 stars for NGC 6266. Thus, the number of extratidal stars found in the previous sections are roughly consistent with a significance of $\sim$19.6 sigma for NGC 6397 and $\sim$11.3 sigma for NGC 2808. Hence, the number of such stars we obtain for NGC 6397 and NGC 2808 are statistically significant. The adopted procedure to select extratidal stars failed to give reliable results in terms of the contamination fraction with regard to the case of NGC 6266 because this cluster has a similar mean PM as that of the field stars. Therefore, we need additional criteria to deal with such clusters.

\subsection{Stricter cuts for NGC 6266}
\label{sec_6266}

To refine our selection of extratidal stars, we imposed stricter cuts on PMs and the difference in color and magnitude with respect to the isochrone of NGC 6266. We selected only those stars whose PMs are matched with the cluster's mean PM within the range of 0.5 times the dispersion in the cluster PM plus the individual error in the PM of the star, that is, \{($\mu_{\alpha}\cos{\delta}\pm 0.5\sigma$), ($\mu_{\delta}\pm 0.5\sigma$)\}$^{star}$ $\lesssim$ \{($\mu_{\alpha}\cos{\delta}\pm 0.5\sigma$), ($\mu_{\delta}\pm 0.5\sigma$)\}$^{cluster}$, where $\sigma_{star}$ and $\sigma_{cluster}$ are the error in the PM measurement of the star and the dispersion in the PM distribution of NGC 6266, respectively. We found 12693 stars matching this criterion. Next, we selected the stars which lie on the annular disk with its center as the center of the cluster and inner and outer radii as the $r_t$ and 5$r_t$ of the cluster, respectively, obtaining 1729 stars. From these stars, only 107 candidates are 0.001 magnitude/color away from the isochrone. We applied the same criteria to the Galactic stars obtained using BGM and found 30 field stars. Hence, we got a $\sim$11.50 sigma detection for this cluster. The position of the extratidal stars on the cluster CMD and their spatial distribution are shown in the bottom panels of Figure~\ref{fig:f1} and Figure~\ref{fig:f2}, respectively. Out of these 107 stars found there, 50 stars are out of the $r_J$ \citep[24.1 arcmin, ][]{deboer19}.
\\

\subsection{Background contamination based on \textit{Gaia} DR2 data}

We also examined the \textit{Gaia} DR2 sources in different adjacent fields to estimate the number of field stars around each cluster. Our motivation is to have an independent, data-driven approach in the estimate of the contamination from field MW stars for the number of extratidal stars we found. Four random fields, having an area of 5 r$_t$  and separated by at least 3.0 deg from the cluster center, in four different regions around each cluster, were searched for extratidal stars using the same criteria as used in Section 2. The adjacent fields have different stellar densities compared to the cluster region. Therefore, to get an estimate of the expected contamination around the cluster, we scaled the number of selected field stars taking the density of the region into account. Table~\ref{tab:extra-tidal} lists the coordinates of the four fields around each cluster, the number of field stars recovered after our selection criteria was applied, the total number of stars in that area, the scaled number of field stars, and the significance of our detections.
\\

According to our estimates, for all the clusters,  there are at least two regions around each of them that have significance values that indicate the number of selected extratidal stars are significant over the number of field stars in that region. For NGC 6397, in the field centered at (RA, Dec) = (270, -54) deg, we found a higher number of stars than the number of extratidal stars found around the cluster itself. This region is located in the trailing side of the cluster, aligned with its orbit. Looking at the density map (Figure~\ref{fig:f2}) and its extratidal stars, overdensities (yellow regions) can be seen along the trailing and leading sides of its orbit. Hence, the overdensity of field stars in this particular region can be attributed, rather, to extended tidal debris from NGC 6397 along its trailing side. This kind of alignment is typical due to bulge or disk shocking \citep{Montuori07}. 

For NGC~2808, our analysis shows a low contamination level for three of the four regions analyzed, but very high contamination towards the eastern region of the cluster. The density map (Figure~\ref{fig:f2}) of the cluster also shows an overdensity of extratidal stars towards that region, hence, this high value of contamination may again be due to  extended tidal debris from the cluster. In the case of NGC~6266, we found two regions having a large fraction of selected field stars, larger than our selection of extratidal stars around the cluster: the field centered at (RA, Dec.) = (255, -25) deg, on the trailing side, and the field centered at (RA, Dec.) = (250, -30) deg, on the eastern side of the cluster. According to the density map for this cluster (bottom panel in Figure~\ref{fig:f2}), there is a significant overdensity of extratidal stars along its past orbit, however there is no such overdensity in the eastern direction. Hence, the large number of field stars found in the trailing side of the cluster can be due to more extended tidal debris, as in the case of NGC 2808, whereas for the other region, this may be due to a high number of MW field stars aligned with the direction towards the Galactic center. Therefore, depending on the direction and the field that is chosen, we find that the number of extratidal stars we recover is significant over the number of field stars in some fields for the three clusters. In other regions, the number of field stars is higher and this could be due to extended tidal debris that is further away from the area studied in this work.
\\

\begin{table*}
        \centering
        \caption{\textit{Gaia} DR2 field stars that pass our selection criteria in 4 different regions around each cluster.}
        \label{tab:extra-tidal}
        \begin{adjustbox}{width=2.0\columnwidth,center}
        \begin{tabular}{|ccccc|} % four columns, alignment for each
\hline
Center (RA, Dec) & Number of field stars &  Total number of stars in the region & Scaled number of field stars$^!$ & Significance$^\#$ \\
\hline
\hline
NGC~6397 & 120 & 1789728 &  &\\
\hline
265, -35 & 15 & 2497841 & 11 & 32.9\\
265, -70 & 34 & 617114  & 99 & 14.7\\
270, -54 & 82 & 610438 & 241 & High contamination\\
250, -54 & 16 & 3789764 & 8 & 39.6\\
\hline
\hline
NGC~2808 & 126 & 101782 &  &\\
\hline
138, -60 & 99 & 183933  & 58   & 8.9\\
138, -70 & 75 & 67880   & 113   & 1.2\\
143, -65 & 77 & 202611  & 39   & 13.9\\
133, -65 & 106 & 62596  & 173   & High contamination\\
\hline
\hline
NGC~6266 & 107 & 159567 &  &\\
\hline
255, -25 & 93 & 103460 & 143 & High contamination\\
255, -35 & 51 & 125075 & 65  & 5.2\\
250, -30 & 72 & 75295 & 152  & High contamination\\
260, -30 & 77 & 144520 & 85  & 2.4\\
\hline
\end{tabular}
\end{adjustbox}
\tablefoot{!-$\text{Scaled number of field stars} (N_{\rm field})=\text{Number of field stars in the region}\times\dfrac{\text{Total number of stars around the cluster}}{\text{Total number of stars in the region}}$ ; \#-$\text{Significance}=(N_{\rm extra-tidal} - N_{\rm field})/\sqrt{N_{\rm field}}$, where $N_{\rm extra-tidal}$ is number of extratidal stars around the cluster.}
\end{table*}

We conclude that the number of potential extratidal star candidates around NGC 6397 and NGC 2808 includes a low number of contaminants from the Galactic population and, therefore, our detection of extratidal stars around these clusters represent real overdensities over the field population. However, we find that the expected number of MW stars toward NGC 6266 is larger given its location towards the dense region of the Galactic bulge, suggesting that our potential extratidal stars can be composed of a mixture of truly extratidal cluster stars and stars from this Galactic population. Future spectroscopic follow-up observations will help to identify the truly members of NGC 6266 according to the radial velocities and spectroscopic metallicities of the individual candidates.

\subsection{Extra-tidal RR Lyrae variable stars around NGC 6266}
\label{RRL}

The field around NGC 6266 has a very high contamination from field stars, hence, we decided to further study its extratidal region of the cluster using RR Lyrae stars from the Optical Gravitational Lensing Experiment (OGLE) database. The RR Lyrae variable stars are excellent standard candles and tracers of extratidal debris around GCs. Recently, \citet{ogle2019} published their new catalog\footnote{\url{ftp://ftp.astrouw.edu.pl/ogle/ogle4/OCVS/blg/rrlyr/}} of 78350 RR Lyrae variable stars, including some variables in the region around NGC 6266. We applied the same criteria as discussed above, but we  selected the ab-type RR Lyrae stars whose PM matches with the cluster within three sigma of the cluster mean PM. We relaxed this cut to include more candidates and get rid of possible contaminants based on the CMD, as RR Lyrae stars should be located at the horizontal branch level of the cluster's CMD. After applying the PM, $r_t$, and CMD criteria, we are left with 11 extratidal RR Lyrae variable stars. The position of these extratidal RR Lyrae stars on the extinction-corrected CMD is shown in Figure~\ref{fig:f1} in pink and the spatial-distribution of these stars is shown in Figure~\ref{fig:f2} using stars as symbols. Overall, 9 out of 11 extratidal RR Lyrae stars lie outside the $r_J$ of the cluster. The presence of these extratidal RR Lyrae variable stars and their spatial distribution is another, independent piece of evidence of the existence of extratidal stars around NGC 6266. It is worth mentioning that these RR Lyrae stars are located much further out from the cluster center than the excess of RR Lyrae reported in \citet{dante18}.

\section{Orbits of the clusters}
\label{sec3}

We successfully identified potential extratidal candidates around NGC~6397, NGC~2808, and NGC~6266. To get a general picture about their specific extra-tidal features, we computed the orbits of each cluster. To this purpose, we used the {\texttt{GravPot16}} code\footnote{https://gravpot.utinam.cnrs.fr}, which employs a physical and realistic (as far as possible) "boxy/peanut" bar model of the Galaxy along with other stellar components \citep[see, e.g.,][]{jose2020}. For this study, we have caried out a backwards integration (until 3 Gyr) of an ensemble (one million simulations per cluster) for orbits of each cluster by adopting the same galactic configurations and a Monte Carlo approach, as described in \citet{jose2020}. 
\\

Table~\ref{Table2} lists the input parameters used in our analysis along with the results. This analysis reveals that the three clusters have high eccentric orbits $e = 0.59$ -- $0.94$, with a perigalactic and apogalactic distance between $r_{peri}$ $\sim$ 0.37 -- 0.41 kpc, and $r_{apo}\sim$ 2.88 -- 14.47 kpc, respectively. The clusters exhibit vertical excursions above the Galactic plane not larger than 3.7 kpc, which indicates that the former may be still interacting with the disk and suggesting that there could be some signatures for the presence of extra-tidal material around these clusters as already been noted in other GC near the Galactic plane such as NGC~6535 and NGC~6254 \citep[see, e.g.,][]{Leon00}.
\\ 

Given that the angular momentum is not a conserved quantity in a model with non-axisymmetric components (e.g., a bar structure), we listed both the minimum and maximum z-component of the angular momentum ($L_z$) in Table~\ref{Table2}. Our simulations reveal that the three clusters have prograde ({$L_{z, min}$, $L_{z, max}$}$< {0, 0}$) orbits with respect to the direction of the Galactic rotation. 
\\

Following the same methodology as in \citet[][]{jose2020}, we provide the classification of the GCs into a specific Galactic component. This classification is based on the location of the cluster on the characteristic orbital energy (($E_{max}+E_{min}$)/2) versus the orbital Jacobi constant ($E_J$) plot. The plot is divided into three different regions corresponding to disk population, stellar halo and bulge/bar population \citep[see Fig. 3 in][]{jose2020}. The position of the cluster in this diagram gives us the membership probability corresponding to each Galactic component. Table \ref{Table2} reveals that NGC 6397 shows a high probability (> 90\%) to belong to the thick disk component,  whereas NGC 2808 is characterized by having a very high probability (> 99\%) of belonging to the inner halo component and NGC 6266 is in the boundary between two Galactic components, indicating that this cluster has a similar probability to be part of the bulge or bar ($<48$\%) and inner disk ($\sim51$\%) for the three patterns speeds of the bar. However, \citet{Perez-Villegas2020}, in adopting a more simplified Galactic model for the MW, recently classified NGC~6266 with a high probability ($>97$\%) of belonging to the bulge or bar component, indicating that its orbital configuration strongly depends on the choice of the gravitational potential assumed for the MW and their observational parameters.
\\

It is worth mentioning that our orbital classification can be compared to other Galactic GCs, including both those formed in situ and those formed in different progenitors, which were only accreted later \citep[see, e.g.,][]{Massari19}. Considering their origin can help us to assess the level up to which the possible extratidal star candidates could be contributing to the stellar populations in the inner Galaxy. Based on the ($E_{max}+E_{min}$)/2 and $E_J$, as envisioned by \citet{Moreno2015} and \citet{jose2020}, and the orbital elements of known GCs and their associated origin according to \citet{Massari19}, we can say that NGC~2808 has an orbital energy consistent with GCs associated to the Gaia-Enceladus-Sausage \citep{Massari19}. This association is in agreement with our classification of an inner halo cluster related to merger events experienced by the MW in early epochs. Therefore, the evidence of extratidal material beyond its tidal radius can give us insights in the formation of the inner stellar halo. On the ($E_{max}+E_{min}$)/2 versus $E_J$ diagram, NGC~6397 occupies the loci dominated by GCs in the main disk and NGC~6266 lies in the group of GCs associated to the main bulge, therefore, any further evidence for extratidal material around these two clusters could provide important clues for disentangling the origin of the chemically anomalous stars identified recently towards the bulge and inner disk \citep[see, e.g.,][]{jose16, Schiavon17, Fernandez-Trincado2017c, jose19a, jose19h, jose19c, jose19halo}.

                \begin{sidewaystable*}
        %\begin{table*}
        \begin{tiny}
                %       \begin{center}
                \setlength{\tabcolsep}{0.5mm}  
                \caption{\textit{Lines 1--5:} Basic parameters of the selected GCs. \textit{Lines 6--17:} Main orbital parameters of the GCs analyzed in this study. The numbers inside parentheses indicate the sensitivity of the orbital elements to the different angular velocity of the bar ($\Omega_{\rm bar}$), which we have computed as the standard deviation of the orbital elements when considering three different values for the bar pattern speeds, $\Omega_{\rm bar} = $33, 43, and 53 km s$^{-1}$ kpc$^{-1}$. \textit{Lines 18--22:} Membership probability for the different bar pattern speed adopted.  
                }
                \centering
                \begin{tabular}{|l | c | c | c | c | c | c | c | c | c |}
                        \hline
Cluster Ids& RA & Dec   & $\mu_{\alpha}\cos(\delta)$ & $\mu_{\delta}$ & $RV$ & d & $\Delta \mu_{\alpha}\cos(\delta)$ & $\Delta \mu_{\delta}$ & $ \Delta RV$  \\
& (deg.)        & (deg.) &  (mas yr$^{-1}$)  & (mas yr$^{-1}$)  & (km s$^{-1}$) & (kpc) & (mas yr$^{-1}$)     & (mas yr$^{-1}$)    & (km s$^{-1}$) \\
\hline
\hline
NGC~2808 & 138.01 & $-$64.86 & 1.01 & 0.27   & 103.57 & 9.60 & 0.05 & 0.05 & 0.27 \\
$\dagger$  &  &  & 0.58   & 2.06     & 101.60   & 9.60 & 0.45   & 0.46   & 0.70  \\
$\dagger\dagger$  &  &  & 1.02   &  0.28 &  103.57 & 10.21 & 0.01 & 0.01 & 0.27 \\
NGC~6397  & 265.17 & $-$53.67 & 3.28 & $-$17.60 & 18.51 & 2.30 & 0.04 & 0.04 & 0.08 \\
$\dagger$  &  &  & 3.69    & $-$14.88  &  18.80 &  2.30 &  0.29 & 0.26 & 0.10  \\
$\dagger\dagger$  &  &  & 3.3 & $-$17.60 & 18.51 & 2.44 & 0.01 & 0.01 & 0.08 \\
NGC~6266 & 255.30 & $-$30.11 & $-$5.05 & $-$2.95 & $-$73.98 & 6.80 & 0.06 & 0.06 & 0.67 \\
$ \dagger$ &  &  &  $-$3.50  & $-$0.82  & $-$70.10  & 6.80 & 0.37 & 0.37  & 1.40 \\
$\dagger\dagger$  &  &  & $-$4.99 & $-$2.95 & $-$73.98 & 6.41 & 0.02 & 0.02 & 0.67 \\
$*$                          & 255.31 & $-$30.11 & $-$5.06 &  $-$2.98 & $-$73.49 &  6.41 & 0.07 & 0.07 & 0.70 \\
\hline
\hline
 Cluster Ids & r$_{\rm peri}$                  &   r$_{\rm apo}$                  &  $ eccentricity (e) $                    &      Z$_{\rm max}$                   &          L$_{\rm z,min}$                   &   L$_{\rm z,max}$         &     $E_{J}$                  &    $E_{\rm char}$                          & Orbit        \\
& (kpc) & (kpc) & &  (kpc) & (10$^{1}$ km s$^{-1}$ kpc)  & (10$^{1}$ km s$^{-1}$ kpc)  & (10$^{5}$ km$^{2}$ s$^{2}$) & (10$^{5}$ km$^{2}$ s$^{2}$) &  \\
\hline
\hline
NGC~2808     &  0.42 $\pm$ 0.08  ( 0.04) &  14.48 $\pm$ 0.11 ( 0.03) & 0.94 $\pm$ 0.01 ( 0.005) &  3.00 $\pm$ 0.10 ( 0.07)  &  $-$36.0 $\pm$ 1.5  ( 1.25) &   $-$16.0 $\pm$ 5.0 ( 2.94) &   $-$1.77  $\pm$ 0.010 (0.03) & $-$1.66 $\pm$ 0.02 ( 0.01) & Prograde \\
 $\dagger$                  &      2.27                         & 10.74                             &  0.649                     &   2.39     & ... & ... & ... & ...  &  ... \\            %            Moreno+2014
  $\dagger\dagger$    &      0.97 $\pm$ 0.02       &  14.76 ± 0.13                &  ...                          &      ..        &  ...  &  ...  &  ... &   ... &  ...  \\             %             Baumgardt+2019
\hline
NGC~6397     &  1.98 $\pm$ 0.07  ( 0.51) &   7.77 $\pm$ 0.04 ( 0.59) & 0.59 $\pm$ 0.01 ( 0.10) &  3.73 $\pm$ 0.07 ( 0.12)  & $-$102.0 $\pm$ 1.5  (11.81) &   $-$56.0 $\pm$ 1.5 (15.11) &   $-$2.28  $\pm$  0.004 (0.06) &  $-$1.97 $\pm$  0.002 (0.05) & Prograde \\
$\dagger$                 &      2.53                         &   5.12                            &  0.34                        &   1.46       &  ...  &  ...  &  ... &   ... &  ...   \\             %        Moreno+2014
  $\dagger\dagger$    &      2.63 $\pm$ 0.03     &    6.23 $\pm$ 0.02          &      ,,,                        &   ...          &  ...  &  ...  &  ... &   ... &  ...  \\             %        Baumgardt+2019
\hline
NGC~6266       &  0.38 $\pm$ 0.17  ( 0.08) &   2.88 $\pm$ 0.14 ( 0.04) & 0.76 $\pm$ 0.09 ( 0.04) &  1.01 $\pm$ 0.14 ( 0.04)  &  $-$43.0 $\pm$ 2.0  ( 1.25) &    $-$9.0 $\pm$ 4.0 ( 2.16) &   $-$2.62  $\pm$  0.01 (0.02) &  $-$2.48 $\pm$ 0.02 (0.01) & Prograde  \\
    $\dagger$             &     1.52                           & 2.63                           & 0.28                       & 0.83            &  ...  &  ...  &  ... &   ... &  ...   \\        %        Moreno+2014
    $\dagger\dagger$ &     0.83 $\pm$ 0.07       &  2.36 $\pm$ 0.09      &   ...                           &      ...            &  ...  &  ...  &  ... &   ... &  ...   \\        %          Baumgardt+2019 
       $ *$                    &     0.35 $\pm$ 0.16         &  2.82 $\pm$ 0.16       & 0.79 $\pm$ 0.09        & 1.10 $\pm$ 0.14  &  ...  &  ...  &  ... &   ... &  ...   \\   %     Perez-Villegas+2020
                        \hline
                        \hline
                        $ \Omega_{\rm bar} = $ & $33$ km s$^{-1}$ kpc$^{-1}$ & $33$ km s$^{-1}$ kpc$^{-1}$ &   $33$ km s$^{-1}$ kpc$^{-1}$ &   $43$ km s$^{-1}$ kpc$^{-1}$  & $43$ km s$^{-1}$ kpc$^{-1}$ &  $43$ km s$^{-1}$ kpc$^{-1}$   &  $53$ km s$^{-1}$ kpc$^{-1}$  & $53$ km s$^{-1}$ kpc$^{-1}$ & $53$ km s$^{-1}$ kpc$^{-1}$  \\
                        Cluster Ids  &  Bulge/Bar & Disk & Stellar Halo  &  Bulge/Bar & Disk & Stellar Halo  & Bulge/Bar & Disk & Stellar Halo \\
                          &   \% & \% &  \%  & \%  & \% &  \% & \% & \% & \% \\
                        \hline
                          NGC~2808 &   0.00      &    0.65   &    99.35 &    0.00    &     1.30  &     98.70 &      0.00  &        2.62 &     97.38 \\
                        NGC~6397 &   0.01      &   90.62   &     9.36  &    0.03    &    95.39  &      4.57  &      0.24  &       97.23 &      2.52 \\
                        NGC~6266   &  45.07      &   54.60   &     0.33 &   47.19    &    52.52  &      0.29 &     48.25  &       51.49 &      0.26 \\
                        \hline 
                \end{tabular} 
        \label{Table2}
        \tablefoot{$\dagger$\citet{Moreno14}; $\dagger\dagger$\citet{Baumgardt2019}; $*$\citet{Perez-Villegas2020}; input parameters employed in this study were taken from \citet{GC}. \citet{Moreno14} and \citet{Baumgardt2019} used $\Omega_{\rm bar}= 55$ km s$^{-1}$ kpc$^{-1}$ and \citet{Perez-Villegas2020} used $\Omega_{\rm bar}=45$ km s$^{-1}$ kpc$^{-1}$.}
        \end{tiny}
\end{sidewaystable*}
%       \end{center}
%\end{table*}

\section{Discussion and  conclusions}
\label{sec5}

We examined the outermost regions of the Galactic GCs NGC~6397, NGC~2808, and NGC~6266 in our search for evidence of extratidal features in the \textit{Gaia} DR2 database. We identified potential extratidal star candidates toward NGC~2808 and NGC~6397, while some possible extratidal signatures seem to be present around NGC~6266. The high reddening (E(B-V)=0.47 mag, H96) and high field density, along with its comparable PM to foreground and background Galactic stars make it difficult to identify extratidal features around this cluster with a high level of confidence. This study yields 120, 126, and 107 extratidal candidates associated to NGC~6397, NGC~2808, and NGC~6266, respectively. These extratidal stars are statistically significant over the field stars in that region of the sky. Our results for each cluster are summarized as follows:
\\

{\bf NGC~6397:} Our result seems to be in good agreement with the results of \citet{Leon00}, where tidal tails for the cluster were reported, although the dust extinction prevented the authors from further exploring their distribution and extent. The cluster has a relatively high eccentric ($e>0.59$) prograde orbit with vertical excursions above the Galactic plane not larger than $3.73$ kpc with a very likely orbit confined to the disk population, which is crossing the Galactic plane every $\sim$0.12 Gyr. Then, the extratidal star candidates could be the effect of the shocks experienced by the cluster with the disk in a short timescale. The extratidal stars for the cluster are asymmetrically distributed around its orbit (see top panel in Figure~\ref{fig:f2}), forming a spiral-like structure from the north-east to south-west direction. These spiral arms can be seen in both the leading and trailing regions around the cluster, resembling the S-shape structured that is considered a characteristic feature of tidal disruption \citep{ray17,ray19}. Moreover, we found a high density of stars satisfying our selection criteria along the direction of the past orbit of the cluster. Based on the shape of the extratidal stars in the cluster's vicinity they can be due to tidal disruption. The cluster's orbit and the high density of stars in the direction opposite to the cluster motion indicate that the nature of a few of the stars can be the result of the disk shocking. Hence, the features and overdensities around the cluster can be the result of a combined effect of tidal disruption and disk shocking.
\\

{\bf NGC~2808:} Most of the stars selected in our study clearly lie on or near the prominent sub-giant branch and horizontal branch of the cluster. Our dynamical analysis shows that this is a cluster that lies in a halo-like orbit, therefore, any signature for extratidal features could help improve our understanding of the origin of stellar properties and the content of the inner halo of the MW. We also find that NGC~2808 is crossing the disk with a frequency of $\sim$0.20 Gyr$^{-1}$, which could explain the asymmetric distribution of the potential extratidal candidates (seen in Figure~\ref{fig:f2}) that exhibit a high stellar density in the trailing region of the cluster with some misalignment with respect to its orbit, which is in good agreement with previous works \citep[see, e.g.,][]{julio17}. Recently, \citet{Sollima20} studied the presence of tidal tails around several GCs, including NGC 2808. The author did not find any coherent tidal tail structure for this cluster beyond 1.5 times the r$_J$. Our study covers up to 1.4 times the r$_J$ (a radius of 1.5 degree from the cluster center) and, therefore, our findings cannot be directly compared to the lack of detections in \citet{Sollima20}. 
\\

{\bf NGC~6266:} Most of the 107 extratidal candidates identified around NGC~6266 follow the red giant branch (RGB) of the cluster. Contamination analysis of the region reveals that the cluster may have tidal tails in the northern and eastern sides. Similar distribution of the stars was also found by \citet{Chun15}. Eastern overdensity found in our analysis is in the trailing part of the cluster. Also, our dynamical study reveals that NGC~6266 is crossing the Galactic plane every $\sim$0.04 Gyr in the inner Galaxy. The extratidal stars are symmetrically distributed around the cluster, resembling the shape of an extended stellar envelope \citep{kuzma16, kuzma17}. This extended stellar halo is in agreement with the results of \citet{Gieles11}, which places this cluster into the expansion dominated phase, which is the internal relaxation the main mechanism producing extratidal stars. Based on the orbit and the contamination analysis, at larger projected distances from the cluster center, some of the extratidal stars  can be the result of a recent disk shock. Despite a high fraction of contaminants is expected in our final sample, this is the best sample that has so far identified possible extratidal feature and provided a motivation for a future spectroscopic follow-up study to confirm or refute the cluster members, in particular towards the inner Galaxy, where some missing pieces still lack in the understanding of the origin of some unusual stars in the inner stellar halo at the same metallicity of NGC~6266.
\\

\citet{Ernst13} used both observations and simulations to conclude that most of the GCs in the MW under-fill their Roche lobe, presenting a mean ration of $r_t$/$r_J$ of 0.48. This is also the case of NGC 6397, NGC 2808 and NGC 6266, having $r_t$/$r_J$ of 0.59, 0.25 and 0.38, respectively. Therefore, any extratidal star based on the adopted r$_t$ is still bound to the cluster and does not necessarily mean that the cluster is under disruption. To better understand the extratidal stars that are fully detached from the cluster and possible disruption process, we used $r_J$. According to \citet{Kupper10}, most of the stars which lie beyond 50\% of the $r_J$ of a given cluster are energetically unbound while beyond 70\% of the $r_J$ almost all stars are detached from the cluster. {Hence, the stars lying outside the $r_t$ but inside the $r_J$ can be termed as potential escapers and the stars situated outside the $r_J$ are fully detached from the clusters.} It can be seen from the density plots (Figure~\ref{fig:f2}) that the selected extratidal stars are located inside as well as outside  the $r_J$ for all the clusters. For NGC 6397, out of the 120 extratidal stars identified in this work, 84 are outside its r$_J$. Hence, up to 70\% of candidates are fully unbound from the cluster. Similarly, for NGC 2808 and NGC 6266, 25.4\%, and 72.9\% stars outside the $r_J$, respectively. The stars that are outside the $r_J$ are fully detached from the cluster, while the stars inside the $r_J,$ but outside the $r_t$ of each cluster, have a higher probability of being pulled out of the $r_J$ as compared to other stars due to the gravitational field of the Galaxy.
\\

Based on the distribution of extratidal stars, we found that most likely NGC 6397 and NGC 2808 suffer from disk shocks and tidal disruption. This is not completely consistent with the position of these clusters in the survivability diagram of \citet{gnedin97}, where NGC 6397 lies outside the survivability diagram, where the "lucky survivor" GCs reside, while NGC 2808 is at the middle of the diagram, having been able to survive another Hubble time. If the internal relaxation is the main mechanism of disruption, we would expect to see an extended stellar envelope around these clusters, which we may not have recovered due to our cuts in magnitude, in the case of NGC 2808. For NGC 6397, however, we searched for stars up to the main sequence and our results do not support a scenario in which this cluster would be surrounded by a stellar envelope. In the case of NGC 6266, the distribution of extratidal stars is homogeneous and resembles an extended stellar halo around the cluster. This is in agreement with the position of the cluster in the diagram of \citet{gnedin97}, in which this cluster is at the edge of the survivability boundary, being affected by internal relaxation and bulge and disk shockings. Radial velocity measurements are therefore needed to confirm the extratidal stars candidates found around these clusters and to determine the disruption mechanisms that are producing the overdensities recovered in this work.
\\

The three selected clusters lie in regions of the Galaxy characterized by different environments and different extinction values. Our analysis shows that if the cluster stars are well-separated from the field stars in the PM plane, using the basic photometric data, with a small dependence on the part of the parallaxes, we were able to extract possible extratidal stars from all the clusters. Our techniques provide us with the best sample of possible extratidal stars based on basic photometric and astrometric observations. The use of PMs and CMDs minimizes the level of foreground and background contamination in the regions where accurate distances to the stars are not available. We cross-matched our sample of extratidal stars with the galaxy and quasar catalog of \citet{Coryn19}. We did not find any match indicating that our data is free from any obvious background contamination from such sources. Hence, in a future work, we plan to apply the same techniques to most of the Galactic GCs in order to study the 3D spatial distribution of clusters that present evidence of extra-tidal stars, with the aim of shedding light on the gravitational potential of our Galaxy.

\begin{acknowledgements}

We thank the anonymous referee for an useful report that helped to improve this paper. We are grateful to Julio Carballo-Bello for useful discussions about the disruption processes in these three globular clusters. R.~K and D.~M are very grateful for the hospitality of the Vatican Observatory, where this collaboration was started. J.~G.~F-T is supported by FONDECYT No. 3180210. D.~M gratefully acknowledges support provided by the BASAL Center for Astrophysics and Associated Technologies (CATA) through grant AFB 170002, and from project FONDECYT No.1170121. This research has made use of the NASA/IPAC Infrared Science Archive, which is funded by the National Aeronautics and Space Administration and operated by the California Institute of Technology. 

\end{acknowledgements}
%%%%%%%%%%%%%%%%%%%%%%%%%%%%%%%%%%%%%%%%%%%%%%%%%%

%%%%%%%%%%%%%%%%%%%% REFERENCES %%%%%%%%%%%%%%%%%%

% The best way to enter references is to use BibTeX:

%\bibliographystyle{mnras}
%\bibliography{example} % if your bibtex file is called example.bib

% Alternatively you could enter them by hand, like this:
% This method is tedious and prone to error if you have lots of references

\bibliographystyle{aa}
\bibliography{NGC6397} % if your bibtex file is called example.bib
\end{document}